\documentstyle[twoside,fleqn,espcrc2]{article}
\input{psfig}

\newcommand{\AmS}{{\protect\the\textfont2
  A\kern-.1667em\lower.5ex\hbox{M}\kern-.125emS}}

\title{The Staggered $\eta'$ with Smeared Operators}
\author{${\rm {L.\,\,Venkataraman}^a}$,\, G. Kilcup\address{Department of Physics,
        The Ohio State University, 174 West 18th Ave, Columbus OH \, 43210}
        and J. Grandy\address{Lawrence Livermore National Laboratory, Livermore CA \, 94550}}
\begin{document}
\begin{abstract}

We present a refined calculation of the $\eta'$ mass
using staggered fermions and Wuppertal smeared operators
to suppress excited state contributions.
We use quenched and dynamical configurations of size
$16^3\times32$, with $N_f=0$, $N_f=2$ and $N_f=4$, and compare our
results with the expected theoretical forms from
quenched, partially quenched, and unquenched chiral
perturbation theory.
\end{abstract}
\maketitle

\section{INTRODUCTION}

The pseudoscalar spectrum of QCD consists of an octet of
mesons which are approximate Goldstone bosons of spontaneously 
broken $SU(3)$ axial flavor symmetry, plus an anomalously
heavy flavor singlet meson, the $\eta'$.
The heaviness of the $\eta'$ is attributed to the effects
of topology~\cite{REF1}. In an $SU(3)$ symmetric world,
$m_{\eta'}$ would obey \begin{equation}
m_{\eta'}^2 = m_0^2 + m_8^2
\end{equation}
where $m_8^2$ is the average mass squared of the octet mesons
which vanishes in the chiral limit, while $m_0^2$ is the 
topological contribution which does not vanish in the chiral limit.
Neglecting $\eta$-$\eta'$ mixing, one plugs in the known meson
masses to obtain the ``experimental'' value
$m_0(N_f=3)=860{\rm MeV}$.

\section{LATTICE CALCULATION OF $m_{\eta'}$}
Extraction of $m_{\eta'}$ from first principles has received 
much attention~\cite{REF4,REF5,REF6,REF7}.
The focus in all these studies has been to calculate the ratio
$R(t)$ defined below
\begin{equation}
R(t) = \frac{\langle \eta'(t)\eta'(0){\rangle}_{2-loop}}{\langle \eta'(t)\eta'(0){\rangle}_{1-loop}}
\label{eqn2}
\end{equation}
where $\eta'(t)$ is the operator that creates or destroys an $\eta'$ 
meson (in terms of quark fields, for staggered fermions, this becomes
$\overline{Q}{\gamma}_5\otimes IQ$). Two point correlation 
function of this operator yields both a disconnected diagram (referred
to as 2-loop in eqn~\ref{eqn2}) and a connected diagram (1-loop).

On dynamical configurations, when $m_{val} = m_{dyn}$,
$R(t)$ takes the following asymptotic form
\begin{equation}
R(t) = \frac{N_v}{N_f}\,\,\lbrack 1\,\, - \,\,B \exp(-\Delta mt)\rbrack
\label{eqn3}
\end{equation}
where $N_v$ is the number of valence fermions, $N_f$ is the number
of dynamical fermions, B is a constant and $\Delta m = m_{\eta'}-m_8$.

For the quenched configurations, infinite iteration of the
basic double pole vertex  does not exist and it
can be shown that the ratio is a linear function of time~\cite{REF4}.
\begin{equation}
R(t) = const. + \frac{m_0^2}{2m_8}\,t
\label{eqn5}
\end{equation}

This style of calculation was employed by the authors of~\cite{REF4}
who obtained a result of $m_0(N_f=3)=751(39){\rm MeV}$ using
quenched configurations and Wilson fermions.  We used staggered
fermions and both dynamical and quenched configurations and
reported a value of $m_0(N_f=3)=730(250){\rm MeV}$ extracted from
dynamical configurations in~\cite{REF5}. 

\section{SIMULATION DETAILS}

The parameters of the ensemble used in the simulation are
shown in table~\ref{TAB1}.  For all of the configurations listed
in table~\ref{TAB1} the inverse lattice spacing is about 2 GeV as obtained
from $m_{\rho}$~\cite{CHEN}. $m_{val}$ for the quenched configurations
has been chosen 10\% higher than that corresponding to dynamical
($m_{dyn}=0.01$, $N_f=2$) so that $m_8$ is same for both.
Propagators were computed using conjugate gradient on the 
128 node OSC Cray T3D machine. For details concerning performance, 
the type of source, the method adopted for calculating the 
disconnected propagator etc., the reader is referred to ~\cite{REF5}.
\begin{table}[tbh]
\caption{The Statistical Ensemble}
\hspace{5pt}
\label{TAB1}
\begin{tabular}{|l|llll|}
\hline
$N_f$  & $m_{dyn}$ & $\beta$  & $N_{samp}$ & $m_{val}$ \\
\hline
0      &  $\infty$ & 6.0      &  83          &  0.011     \\
       &           &          &              &  0.022     \\
       &           &          &              &  0.033     \\
2      &  0.01     & 5.7      &  79          &  0.01      \\
       &           &          &              &  0.02      \\
       &           &          &              &  0.03      \\
2      & 0.015     & 5.7      &  50          &  0.01      \\
       &           &          &              &  0.015     \\
2      & 0.025     & 5.7      &  34          &  0.01     \\
       &           &          &              &  0.025    \\
4      & 0.01      & 5.4      &  70          &  0.01     \\
\hline
\end{tabular}  
\vspace{-0.7cm}
\end{table}

\subsection{Smearing}

The disconnected data is noisy and clean signals exist only for the first
few time slices. Smearing helps in reducing excited state
contributions from the first few time slices, thus enabling a 
reliable extraction of $\Delta m$. We use the gauge invariant Wuppertal
smearing technique.
\begin{figure}[hbt]
\vspace{-0.2cm}
\centerline{\psfig{figure=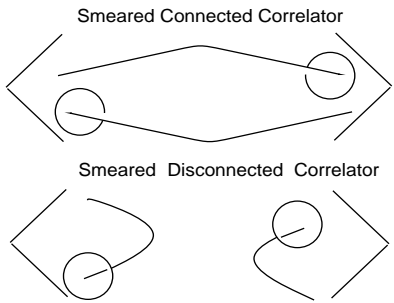,height=4.0cm,width=6.0cm}}
\vspace{-0.7cm}
\caption{Correlators from Smeared Operators}
\label{figcorr}
\vspace{-0.2cm}
\end{figure}
For the connected contraction we computed two propagators, 
one with a smeared source and  point-like sink and the other
with a point-like source and  smeared sink (see Fig.~\ref{figcorr}).
Correspondingly, for the disconnected loops, the sink end was 
smeared in the same way(SS).
\begin{figure}[hbt]
\vspace{-0.2cm}
\centerline{\psfig{figure=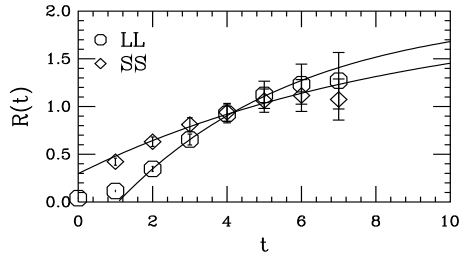,height=3.3cm,width=6.0cm}}
\vspace{-0.8cm}
\caption{Ratio from SS and LL correlators at $m_{val}=m_{dyn}=0.01$}
\label{fig1}
\vspace{-0.4cm}
\end{figure}

Fig.~\ref{fig1} compares the ratio plot with and without 
smearing on dynamical configurations. In the initial few time
slices both the data are different but they begin to
coincide after a few time slices as they should.

\section{RESULTS}

\begin{figure}[h]
\vspace{-0.3cm}
\centerline{\psfig{figure=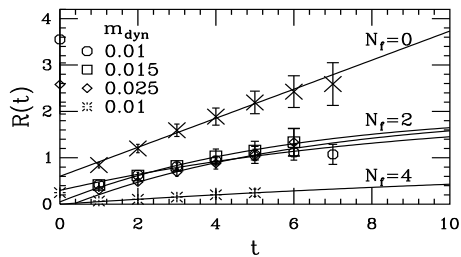,height=3.3cm,width=6.0cm}}
\vspace{-0.8cm}
\caption{$N_f$ dependence of ratio}
\label{fig2}
\vspace{-0.5cm}
\end{figure}

Fig.~\ref{fig2} shows the form of the ratio on all the 
configurations listed in Table~\ref{TAB1}.
For $N_f=2$ and $N_f=4$, the points shown are for $m_{val} = m_{dyn}$,
while for $N_f=0$ we plot $m_{val}=0.011$.
It is gratifying to see that this observable clearly distinguishes
the number of flavors.

\begin{figure}[hbt]
\vspace{-0.2cm}
\centerline{\psfig{figure=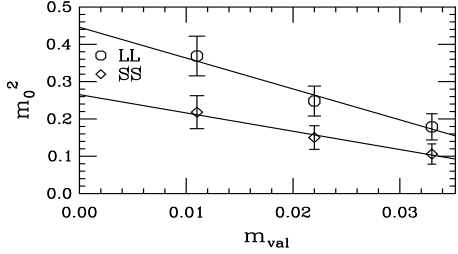,height=3.3cm,width=6.0cm}}
\vspace{-0.6cm}
\caption{Chiral extrapolation of $m_0^2$}
\label{fig3}
\vspace{-0.7cm}
\end{figure}

One extracts $m_0^2$ from a linear fit to the quenched data.
Fig.~\ref{fig3} shows $m_0^2$ versus $m_{val}$ obtained from both 
local and the smeared data, fit linearly and extrapolated to
the chiral limit. As is expected, $m_0^2$ does not vanish in the chiral limit.

\begin{figure}[hbt]
\vspace{-0.2cm}
\centerline{\psfig{figure=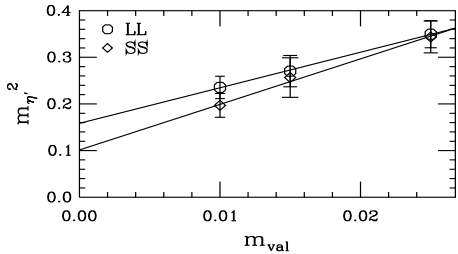,height=3.3cm,width=6.0cm}}
\vspace{-0.6cm}
\caption{Chiral extrapolation of $m_{\eta'}^2$}
\label{fig4}
\vspace{-0.7cm}
\end{figure}

The $N_f=2$ data shown in Fig.~\ref{fig2} is fit to the form of eqn~\ref{eqn3}.
From the fit one extracts $\Delta m$ and hence $m_{\eta'}$ for all
the values of $m_{dyn}$ shown.  It can be seen in Fig.~\ref{fig4} 
that $m_{\eta'}$ does not vanish in the chiral limit.

When $m_{val} \ne m_{dyn}$, the ratio $R(t)$ takes the form 
\begin{equation}
R(t)  = \,\,\frac{N_v}{N_f}\,\lbrack\, At\,\, -\,\, B\exp(-\Delta mt)\,\, + \,\,C\rbrack
\label{eqn4}
\end{equation}
Our data are not precise enough to allow a four parameter fit,
but using lowest order $PQ\chi PT$ (and neglecting the 
small momentum dependent self-interaction $\alpha$) we can
express $A$, $B$ and $C$ in terms of one unknown parameter,
$m_0^2$.
For $m_{val} > m_{dyn}$ ($N_f=2, m_{dyn}=0.01$) data we 
get reasonable fits and find a partially quenched result of
$m_{\eta'}(N_f=3)=876 \pm 16$ MeV,
remarkably consistent with the fully quenched and fully dynamical
data.
For $m_{val} \le m_{dyn}$ the $\chi^2$ is not reasonable.
On the other hand, one need not venture far from lowest
order $PQ\chi PT$ to fit the data.  For example, from two
parameter fits to the dynamical data, one obtains $Z'/Z$,
the ratio of the residues for creating $\eta$ and $\eta'$.
While this ratio is set to unity at lowest order, the
data prefer values 20-30\% larger, an amount which could
easily be accommodated by
higher order chiral and $O(1/N_c)$ corrections.
\begin{figure}[h]
\vspace{-0.3cm}
\centerline{\psfig{figure=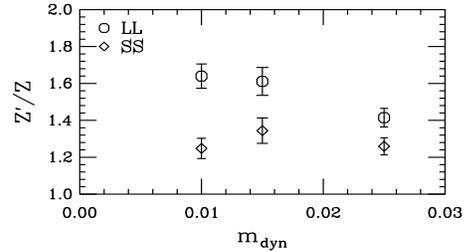,height=3.3cm,width=6.0cm}}
\vspace{-0.6cm}
\caption{$Z'/Z$ versus $m_{dyn}$}
\label{fig5}
\vspace{-0.5cm}
\end{figure}

\section{CONCLUSIONS}

The value, we obtain for $m_{\eta'}$ in the chiral limit are summarized
in the table below. Smeared operators have produced a lower value 
for $m_{\eta'}$  than that obtained from local operators.
\begin{table}[h]
\caption{$m_{\eta'}(N_f=3)$ in the chiral limit}
\label{TAB2}
\begin{tabular}{|ccc|}
\hline
        &Quenched     &Dynamical ($m_{dyn}=m_{val}$)\\
        &(MeV)        &(MeV)   \\
\hline
LL & 1156(95)   & 974(133) \\
SS & 891(101)   & 780(187) \\
\hline
\end{tabular}
\vspace{-0.4cm}
\end{table}
Within the quoted statistical errors, the values obtained above 
are consistent with  experiment.

\end{document}